\documentclass{moriond}

\def\be{\begin{equation}}
\def\ee{\end{equation}}
\def\bea{\begin{eqnarray}}
\def\eea{\end{eqnarray}}

\newcommand{\Photo}{}

\begin{document}
\vspace*{4cm}
\title{Status of the KM3NeT real-time analysis framework}
\author{Martina Marconi, on behalf of the KM3NeT Collaboration}
\address{Università di Genova - Dipartimento di fisica, INFN – Sezione di Genova,\\
Via Dodecaneso 33, Genova, 16146 Italy}

\maketitle
\abstracts{Multi-messenger astronomy requires real-time systems capable of rapidly responding to external alerts and sharing significant detections with partner observatories. KM3NeT, a deep-sea Cherenkov neutrino telescope in the Mediterranean Sea, is actively contributing to these efforts through a dedicated real-time analysis framework. It comprises two detectors - ARCA, optimised for TeV–PeV neutrinos, and ORCA, for GeV–TeV neutrinos - both also sensitive to MeV neutrinos from core-collapse supernovae, providing a wide field of view and an almost continuous duty cycle. The framework performs low-latency event reconstruction and classification, follows up external alerts from the multi-messenger community, monitors for core-collapse supernova neutrino bursts, and autonomously identifies and distribute cosmic neutrino alerts. Now in advanced commissioning, the KM3NeT real-time alert system represents a major step toward rapid, coordinated multi-messenger observations.}

\section{Real-time multi-messenger astronomy with KM3NeT}
Multi-messenger astronomy combines observations of cosmic sources through photons, cosmic rays, gravitational waves, and neutrinos. Coordinated measurements across these channels are crucial for gaining a comprehensive understanding of the most energetic phenomena in the Universe, since each messenger probes different aspects of a source. Furthermore, many astrophysical objects are transient or fast-fading, making low-latency detection and prompt follow-ups essential. Instruments like KM3NeT, with wide fields of view, nearly continuous duty cycles, and real-time analysis capabilities, enable rapid response to external alerts and can issue alerts for significant detections, supporting the global multi-messenger effort.

\subsection{The KM3NeT neutrino telescope}
KM3NeT (Cubic Kilometre Neutrino Telescope) is a deep-sea Cherenkov neutrino observatory under construction in the Mediterranean Sea, currently taking data in partial configuration~\cite{KM3Net:2016zxf}. It detects Cherenkov light induced by charged particles produced by neutrino interactions in seawater using a three-dimensional array of Digital Optical Modules (DOMs), each equipped with 31 3-inch photomultiplier tubes (PMTs). The DOMs are arranged vertically along flexible lines, called detection units (DUs), which are anchored to the seafloor; each DU hosts 18 DOMs. KM3NeT comprises two complementary detectors: KM3NeT/ARCA, optimized for TeV–PeV neutrinos and currently with 51 deployed detection units; and KM3NeT/ORCA, optimized for GeV–TeV neutrinos and currently with 33 deployed detection units. Both detectors are also sensitive to MeV-scale neutrinos through an increase in coincident PMT signals within single DOMs across the entire detector.

\subsection{The KM3NeT real-time analysis framework}
A real-time analysis framework has been developed by the KM3NeT Collaboration~\cite{Mastrodicasa:2025Nr}. Data from both ARCA and ORCA are continuously mirrored to dedicated dispatchers, which perform fast low-latency event reconstruction and classification. Events are reconstructed under both track-like and shower-like hypotheses, and machine-learning-based algorithms are applied to suppress the dominant atmospheric muon background, with a total processing latency below $\sim$10\,s for ARCA and $\sim$15\,s for ORCA. Reconstructed events are sent to a central dispatcher coordinating external alert follow-ups, core-collapse supernova detection, and the selection of and dissemination of neutrino alerts candidates.

\section{Follow-ups of external alerts}
\begin{figure}[h]
    \centering
    \includegraphics[width=1.0\linewidth]{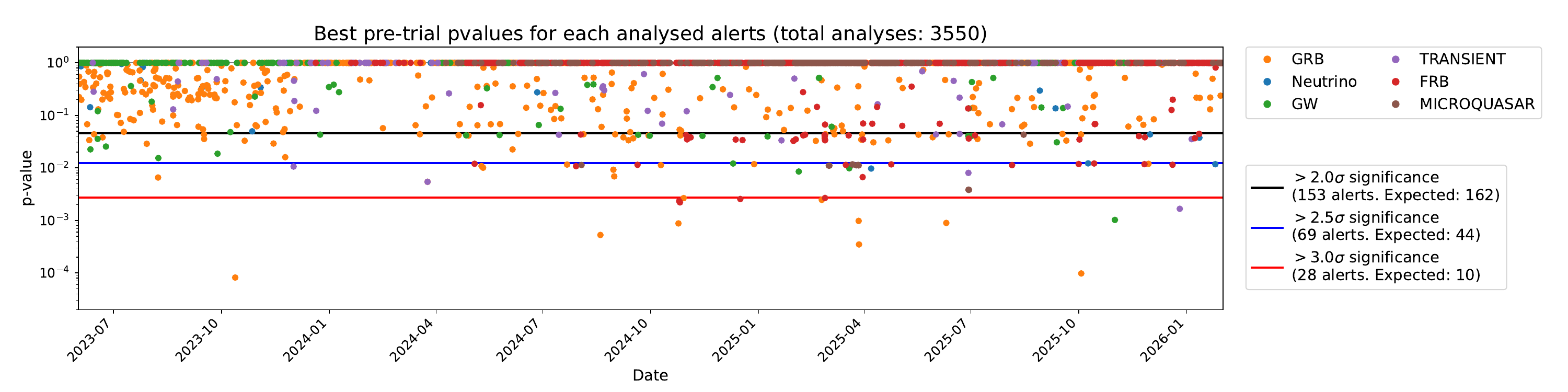}
    \caption{Timeline of the best pre-trial $p$-values computed for follow-ups performed with data from both KM3NeT/ARCA and KM3NeT/ORCA. Horizontal lines indicate the $2\sigma$, $2.5\sigma$, and $3\sigma$ significance thresholds.}
\label{fig:pvalues}
\end{figure}

Since June 2023, the KM3NeT real-time system has analysed over 3550 external alerts. External triggers from internal and external brokers are classified as gamma-ray bursts (GRBs), gravitational wave (GW) events, neutrino alerts, fast radio bursts (FRBs), general transients and microquasar flares, and processed with an automatic space-time correlation analysis~\cite{Filippini:2025QT}.\\
The search for a neutrino counterpart uses a binned ON/OFF technique~\cite{KM3NeT:2024nwb}. Events in the signal region (ON) - defined by a space-time window around the external alert - are compared to that in control regions (OFF) with similar background conditions but no expected signal, i.e. elevation bands that follow the local movement of the ON region due to Earth's rotation. Multiple time windows are tested depending on the alert type, and selection cuts are optimized as a function of elevation to suppress background while controlling statistical uncertainties. Results are expressed as pre-trial $p$-values; in the absence of a signal, upper limits are computed for spectral indices $\gamma = 2.0,\,2.5,\,3.0$.\\
Currently, only track-like KM3NeT events are used; the inclusion of shower-like events is ongoing. As shown in Figure~\ref{fig:pvalues}, no significant correlations have been observed, and the number of analyses exceeding the 2$\sigma$, 2.5$\sigma$, and 3$\sigma$ thresholds is consistent with background expectations. Public reporting of results will begin soon.

\section{Core-Collapse Supernovae real-time detection}
Core-collapse supernovae (CCSNe) emit a burst of $\sim$10\,MeV neutrinos in less than 0.5\,s, requiring fast automated detection. KM3NeT detects MeV electron anti-neutrinos via inverse beta decay. As positron tracks are too short for single-event reconstruction, detection relies on excesses of hit coincidences in single DOMs, evaluated every 100 ms over 500 ms windows and combined across detectors to compute joint significance and false alert rates.\\
Current detector configurations are sensitive to the vast majority of potential CCSN candidates in the Milky Way, and the full-configuration detectors (ARCA230 and ORCA115) will achieve complete Galactic sensitivity across a wide range of progenitor masses~\cite{ElHedri:2025Yv}.\\
Within the KM3NeT real-time analysis framework, three complementary pipelines are in place: (1) a \textbf{real-time} pipeline that continuously monitors the coincidence level and sends alerts to the SNEWS~2.0 (Supernova Early Warning System)~\cite{SNEWS:2020tbu} network upon detection of a significant excess; (2) a \textbf{quasi-online} pipeline, currently in testing, that performs more detailed analyses with accumulated statistics, aiming to reconstruct the neutrino light curve and determine the neutrino arrival time, which is reported to the SNEWS~2.0 timing tier; and (3) a \textbf{triggered follow-up} pipeline that re-analyses archival data in response to external CCSN  triggers.

\section{Neutrino alerts sending system}
\begin{figure}[h]
\centering
\begin{minipage}[t]{0.48\linewidth} 
    \centering
    \includegraphics[width=\linewidth]{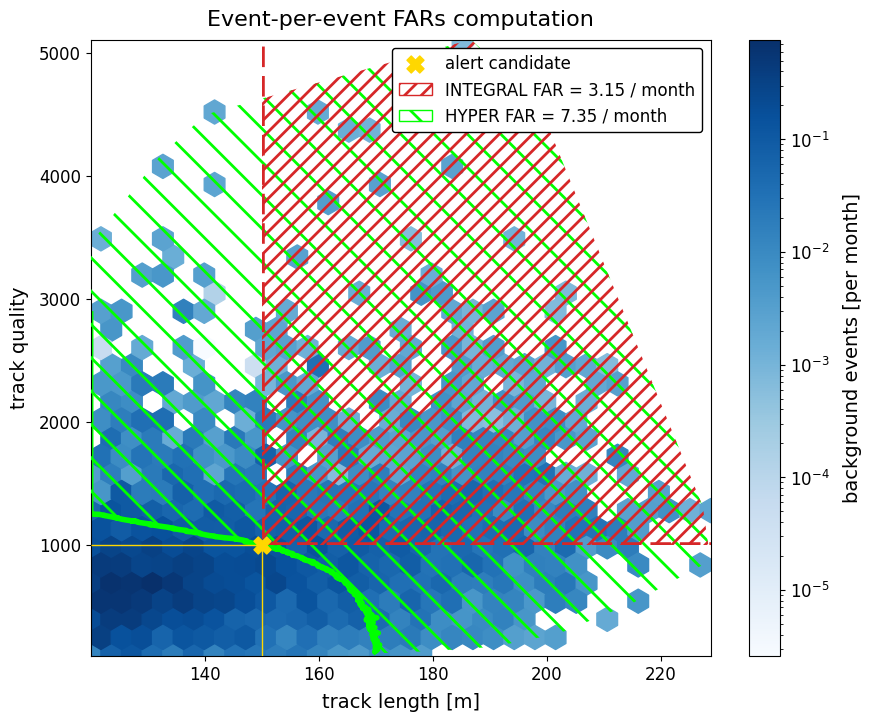}
    \caption{Illustration of the event-per-event FAR computation for a high-energy alert candidate (yellow marker) in the track length vs.\ track quality plane. The red hatched region shows the Integral FAR selection surface; the green hatched region shows the corresponding Hyper FAR surface. The background event density (blue hexbins) is derived from Monte Carlo simulations.}
    \label{fig:money}
\end{minipage}
\hfill
\begin{minipage}[t]{0.48\linewidth}
    \centering
    \includegraphics[width=\linewidth]{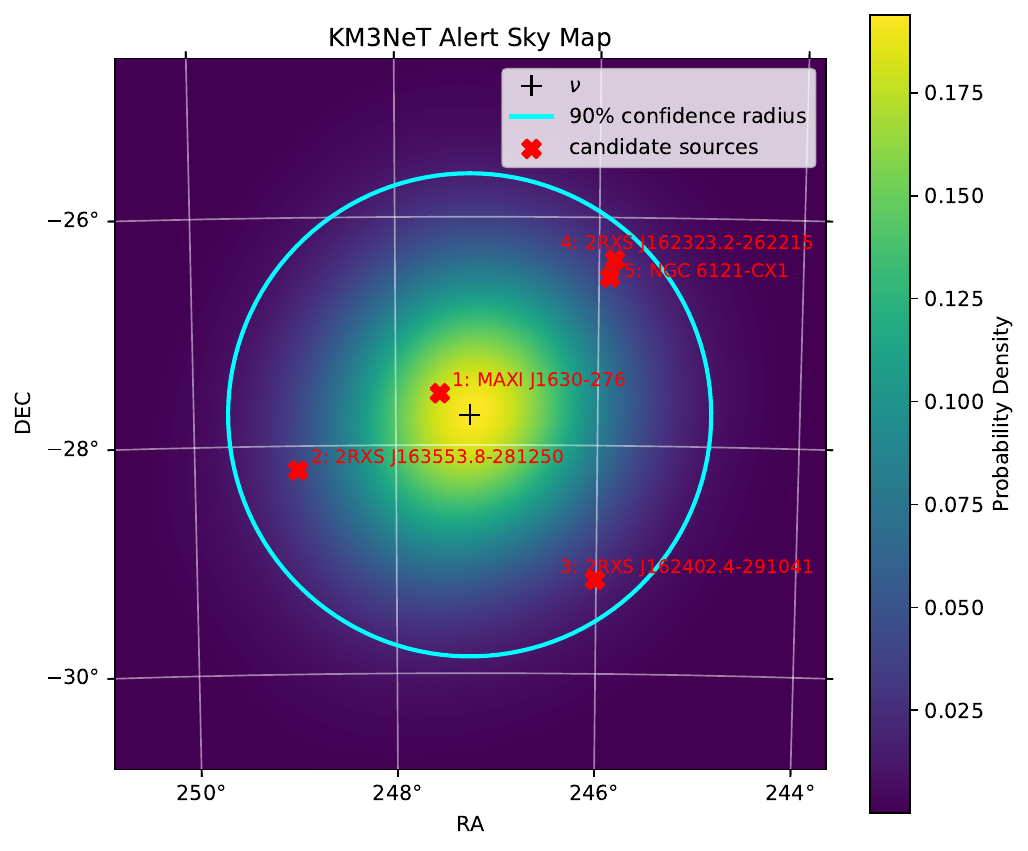}
    \caption{Example KM3NeT alert sky map showing the HEALPix probability density map with the 90\% containment contour (cyan) and ranked candidate astrophysical counterparts (red markers) identified by the astro module within the error region.}
    \label{fig:astro_skymap}
\end{minipage}
\end{figure}

The alert system continuously monitors reconstructed events, autonomously identifying cosmic neutrino candidates and distributing the most significant detections as GCN notices with a target latency of $<$ 3 minutes.\\
Two independent event selection strategies are implemented. The \textbf{high-energy event selection} targets well-reconstructed, upgoing, high-energy track-like events with good angular resolution, likely of cosmic origin, selected on the basis of reconstructed muon energy, track length, and reconstruction quality. The \textbf{multiplet selection} targets pairs of events arriving from compatible sky directions within a short time window, which could indicate a flaring or transient astrophysical source.\\
For each neutrino alert candidate, two complementary False Alert Rates (FARs) are computed. The \textit{Integral FAR} counts the number of Monte Carlo simulated background events with more extreme parameter values than the candidate in the multi-dimensional selection space. However, due to degeneracies in this space, a fixed Integral FAR threshold would result in an uncontrolled alert emission rate. A second quantity, the \textit{Hyper FAR}, is therefore introduced: it is defined as the number of Monte Carlo background events whose Integral FAR is equal to or lower than that of the candidate, effectively reducing the problem to one dimension and providing a direct handle on the physical alert rate. The Hyper FAR is used as the final selection metric and is reported in the GCN notice. This is illustrated in Figure~\ref{fig:money} for a high-energy candidate event.\\
For each selected alert candidate, a per-event sky error region is computed using \textbf{on-the-fly simulations}: $N$ pseudo-events are generated by re-sampling the detected event, accounting for per-event fluctuations, reconstruction effects, detector systematics, and neutrino-muon kinematics. The output is a HEALPix probability sky map and the 50\%, 68\%, and 90\% containment radii, produced with a latency of 1.5–2 minutes.\\
The \textbf{astro module} cross-correlates the HEALPix map with catalogues of known astrophysical sources (currently the RFC~\cite{Petrov_2025}, 2RXS~\cite{Boller_2016}, 4FGL-DR4~\cite{Fermi-LAT:2022byn,Ballet:2023qzs} and XRBcats~\cite{Neumann:2023xrb,Avakyan:2023xrb} catalogues) to identify potential counterparts and provide a ranked list of candidate sources for electromagnetic follow-up observations, as illustrated in Figure~\ref{fig:astro_skymap}. For each catalogue source, a combined association score is built from three criteria: the HEALPix probability density at the source position, the source brightness relative to other catalogue entries, and the source variability in a given energy band at the time of the neutrino detection. A first version of the module is operational.\\
A final decision module aggregates all results: an alert is published if the event reaches Hyper~FAR~$<$~1/month in any selection pipeline, or if significant astrophysical counterparts are identified even below this threshold, in which case the effective FAR incorporates the astrophysical prior. The expected alert rate from KM3NeT is 1-2 events per month. The resulting GCN  notice\footnote{\url{https://github.com/nasa-gcn/gcn-schema/tree/main/gcn/notices/km3net}} includes detector information, event properties, sky localisation (HEALPix probability sky map and containment radii), and the list of candidate astrophysical counterparts. The full system is in advanced commissioning, with public alert distribution expected by summer 2026.

\section{Conclusion}
The KM3NeT real-time analysis platform is continuously operating on data from both ARCA and ORCA, performing event reconstruction, classification, and multi-messenger analyses. The external alert follow-up system has been operational since June 2023, with more than 3500 alerts analysed and no significant neutrino counterparts observed to date. The CCSN real-time detection pipeline is in place, with the quasi-online analysis currently in testing. The full neutrino alert sending system is in advanced commissioning, with the first notices to partner observatories expected by summer 2026. As the detector grows toward its complete configuration, KM3NeT is set to become an increasingly prominent player in the global real-time multi-messenger network.

\section*{References}
\bibliography{moriond}


\end{document}